\documentstyle[11pt,twoside,jltp,psfig,amssymb]{article}

\title{Resonant Raman Study of Superconducting Gap and Electron-Phonon Coupling in $\bf YbBa_2Cu_3O_{7-\delta}$}

\author{S. Ostertun\address{Institut f\"ur Angewandte Physik, 
Jungiusstra{\ss}e 11, D-20355 Hamburg}, 
J. Kiltz, T. Wolf\address{Forschungszentrum Karlsruhe, ITP, Postfach 3640, D-76021 Karlsruhe}, and A. Bock}

\runninghead{S. Ostertun {\boldmath $et\
al.$}}{Resonant Raman Study of $\bf YbBa_2Cu_3O_7$}
       
\begin{document}
\def\ch#1{$\rm #1$}
\newlength{\myparindent}
\setlength{\myparindent}{\parindent}
\def\Bg{$B_{\rm 1g}$}
\def\cm{$\rm cm^{-1}$}
\begin{abstract}
We investigate the electronic background as well as the 
O2-O3 mode at $\it 330\,cm^{-1}$ of highly doped $YbBa_2Cu_3O_{7-\delta}$ in 
$B_{\rm 1g}$ symmetry. 
Above the critical temperature $T_{\rm c}$ the spectra consist of an almost constant 
electronic background and superimposed phononic excitations.
Below $T_{\rm c}$ the superconducting gap opens and the electronic background redistributes 
exhibiting a $2\Delta$ peak at $\it 320\,cm^{-1}$.
We use a model that allows us to separate the background from the phonon.
In this model the phonon intensity is assigned to the coupling of the phonon to inter- and intraband 
electronic excitations.
For excitation energies between 1.96 eV and 2.71 eV the electronic background exhibits 
hardly any resonance. 
Accordingly, the intraband contribution to the phonon intensity is not affected. 
In contrast, the interband contribution vanishes below $T_{\rm c}$ at 1.96 eV 
while it remains almost unaffected at 2.71 eV.

PACS numbers: 74.25.Gz, 74.25.Kc, 74.62.Dh, 74.72.Bk, 78.30.Er
\end{abstract}
\maketitle
\vspace{0.3in}
The Fano-type line shape of the {\Bg} mode in Raman experiments 
in the $R\rm Ba_2Cu_3O_7$ ($R$-123) system with $R$=rare earth or yttrium 
has been subject of several investigations.\cite{Friedl:90,Chen:93,Devereaux:95,Bock:99}.
Using extended Fano models like those presented by Chen {\em et al.}\cite{Chen:93} and 
Devereaux {\em et al.},\cite{Devereaux:95} 
the self-energy contributions to the phonon parameters as a consequence of 
the interaction of the phonon with low-energy electronic excitations
can in principle be obtained. 
Moreover, a measure of the electron-phonon coupling and the ``bare'' phonon intensity,
i.e., the one resulting from a coupling to interband excitations, can be 
estimated from a detailed analysis of the Raman spectra. 
Therefore, a simultaneous description of the real and the imaginary part of the 
electronic response function $\chi^e(\omega)=R^e(\omega)+i\varrho^e(\omega)$ of the intraband excitations 
is needed.
Such a description has recently been presented by us and applied to Ca- and Pr-doped Y-123-films.\cite{Bock:99}
Here, we will use our description in order to investigate the different contributions to the 
{\Bg} phonon intensity in overdoped $\rm YbBa_2Cu_3O_7$ (Yb-123) and their resonance properties as well as the 
resonance of the pair-breaking peak ($2\Delta$ peak) in {\Bg} symmetry. 

We study a fully oxygenated high-quality Yb-123 single crystal grown with a self flux method\cite{Wolf:89}. 
Due to the high oxygen content and the small rare earth ion radius\cite{Lin:95} the
crystal is overdoped ($T_{\rm c}$=76~K).
$B_{\rm 1g}$ Raman spectra [$\rm z(x',y')\bar{z}$ in Porto notation] have been taken
using laser lines at 458, 514, and 633 nm (2.71, 2.41, and 1.96 eV) in a setup described 
elsewhere.\cite{Ruebhausen:97b} 
They have been corrected for the spectral response of spectrometer and detector. 
For a comparison of the spectra obtained with different excitation energies
the cross-section is calculated from the efficiencies\cite{eff:cross} using ellipsometric data of Y-123.\cite{Kircher:91}.
All given temperatures are actual spot temperatures\cite{Bock:95} with typical heatings between 5 K and 15 K.
In order to describe the line shape of the {\Bg} phonon we subdivide the Raman cross-section $I_{\rm c}(\omega)$
into a sum of the electronic response $\varrho_*(\omega)$ and an electron-phonon interference term $I_{\rm p}(\omega)$:\cite{Bock:99}
\begin{equation}\label{eq:phonon}
I_{\rm p}(\omega)=\frac{C}{\gamma(\omega)\left[1+\epsilon^2(\omega)\right]}\times
\frac{R_{\rm tot}^2(\omega)-2\epsilon(\omega)R_{\rm tot}(\omega)\varrho_*(\omega)-\varrho_*^2(\omega)}{C^2}.
\end{equation}
The constant $C=A\gamma^2/g^2$ is a parameter for the intensity where $\gamma$ represents the 
symmetry element of the electron-phonon vertex projected out by the measurement geometry and 
$g$ is the lowest order expansion coefficient of the electron-phonon vertex describing the 
coupling to non-resonant intraband electronic excitations.
$R_*(\omega)+i\varrho_*(\omega)=Cg^2\chi^e(\omega)$ is the electronic response in the measured units.
While $\varrho_*(\omega)$ can be obtained directly from the spectra, $R_*(\omega)$ has to be calculated via 
a Hilbert transformation.
$R_{\rm tot}(\omega)=R_*(\omega)+R_0$ with $R_0=Cg(g_{\rm pp}/\gamma)$ where $g_{\rm pp}$ is an 
abbreviated ``photon-phonon'' vertex that describes the coupling to resonant interband 
electronic excitations.\cite{Devereaux:95} The renormalized phonon frequency $\omega_{\nu}(\omega)$ and linewidth $\gamma(\omega)$
are given by $\omega_{\nu}^2(\omega)=\omega_{\rm p}^2-2\omega_{\rm p}R_*(\omega)/C$ and 
$\gamma(\omega)=\Gamma+\varrho_*(\omega)/C$, respectively, and 
$\epsilon(\omega)=\left[\omega^2-\omega_{\nu}^2(\omega)\right]/\left[2\omega_{\rm p}\gamma(\omega)\right]$.
The bare phonon intensity $I_{\rm pp}$ resulting from the coupling to interband excitations is given by
$I_{\rm pp}=\frac{\pi}CR_0^2$~(Ref.\,\,4). 
The imaginary part of the measured electronic response (background) is modeled by two contributions:\cite{Bock:99}
$I_{\infty}\tanh (\omega/\omega_{\rm T})$ and $I_{\rm red}(\omega,\omega_{2\Delta},\Gamma_{2\Delta},I_{2\Delta},I_{\rm supp})$.
The first term describes the incoherent background with a crossover frequency $\omega_{\rm T}$ 
and the second the redistribution below $T_{\rm c}$ using two Lorentzians, one is centered at the $2\Delta$ 
peak with the intensity $I_{2\Delta}$ and the other, proportional to $I_{\rm supp}$, describes the 
suppression between $\omega=0$ and $\omega=2\Delta$.

Figure~\ref{fig:spectra} (a) displays the {\Bg} cross-section of Yb-123 at 20 K obtained 
with $\hbar\omega_{\rm i}=2.71$~eV as well as its description.
For the description we use
Eq.~(\ref{eq:phonon}) for the {\Bg} phonon and the Ba mode, 
Lorentzians for all other modes, and the background contributions stated above. 
The description yields a $2\Delta$~peak at $\approx320\rm\,cm^{-1}$.
Assuming that the background is not resonant, a calculated spectrum 
with vanishing bare phonon intensity ($R_0=0$) for the {\Bg} phonon is drawn 
in Fig.~\ref{fig:spectra} (b), where other phonons are dropped for clarity. 
We compare this calculation to the cross-section
obtained with $\hbar\omega_{\rm i}=1.96$~eV and 20 K and find a good agreement.

Results of the analysis of the $B_{\rm 1g}$~phonon line shape for measurements with $\hbar\omega_{\rm i}=2.71$~eV
are shown in Fig.~\ref{fig:frequency}. 
It turns out that the strong broadening as well as the slight softening of the renormalized phonon can 
entirely be assigned to the redistributing background leaving anharmonic 
decays for the bare phonon parameters.
As we obtained similar result with $\hbar\omega_{\rm i}=2.41$~eV we used a
fixed parameter set $\Gamma(T)$, $\omega_{\rm p}(T)$ 
for all excitation energies.
This is especially important for the spectra recorded with 
$\hbar\omega_{\rm i}=1.96$~eV, where 
the decreasing or even vanishing bare phonon intensity hinders a 
reliable 
determination of the phonon parameters.

The upper panels of Fig.~\ref{fig:intensity} display the peak height of the $2\Delta$~peak  
obtained with our description of the electronic background. 
Obviously, the $2\Delta$ peaks vanish above $T_{\rm c}$. 
The remaining peak intensity above $T_{\rm c}$ for $\hbar\omega_i=2.71$\,eV is just
a compensation of a slightly underestimated electron-phonon coupling.
Below $T_{\rm c}$ the intensities increase in a
monotonic fashion saturating, more or less pronounced, at low temperatures. 
With respect to the resonance properties of the $2\Delta$~peak we find a decreasing intensity with decreasing
excitation energy.
This partly explains the discrepancy between 
the calculated and the measured cross-sections 
shown in Fig.~\ref{fig:spectra} (b).
The energy of the $2\Delta$~peak decreases only slightly with 
increasing temperature 
from $\approx$ 310 cm$^{-1}$ at 30 K down to $\approx$ 260 cm$^{-1}$ at 70 K.

Regarding the temperature dependencies of the bare phonon intensity $I_{\rm pp}$
in Fig.\,\ref{fig:intensity} we find similar behavior 
for the data sets obtained
with $\hbar\omega_{\rm i}=$2.71~eV and 2.41~eV. 
They exhibit a slight decrease with decreasing temperature being not
affected by $T_{\rm c}$. 
In contrast, we find a dramatically decreasing intensity with $\hbar\omega_{\rm i}=$1.96~eV
below $T_{\rm c}$.
The overall decrease of $I_{\rm pp}$ for $T>T_{\rm c}$ with decreasing excitation energy is similiar to the behavior of
$I_{2\Delta}$ for $T\to 0$, however, more pronounced.

The sudden drop of $I_{\rm pp}$ below $T_{\rm c}$ for $\hbar\omega_{\rm i}=1.96$~eV suggests
a superconductivity-induced closing of the resonant excitation channel of the phonon. 
For even lower excitation energies $I_{\rm pp}$ vanishes almost completely 
for $T\to 0$ as we have observed with $\hbar\omega_{\rm i}=1.71$~eV and 1.58~eV.
Above $T_{\rm c}$, however, the decreasing intensity with decreasing excitation energy 
appears to continue monotonically.
This suggests that more fundamental changes of the band structure take place below $T_{\rm c}$.
They will most likely appear around 
the van Hove singularity at $(\frac{\pi}a,0)$ where the 
electron-phonon coupling is 
enhanced.\cite{Devereaux:95}
It remains open at present how far the band structure changes inferred from our data are related 
to the anomalies around 2~eV observed 
in thermal-difference reflectance spectroscopy,\cite{Holcomb:96}
or to the missing

%
%
\noindent
spectral weight deduced from a sum-rule type analysis of c-axis optical 
conductivity data.\cite{Basov:99}

The authors thank U. Merkt for encouragement.
S. O. acknowledges a grant of the Deutsche Forschungsgemeinschaft via the Gradiuiertenkolleg 
``Physik nanostrukturierter Festk\"orper''.

\clearpage
\begin{figure}
\caption{}{(a) {\Bg} Raman cross-section $I_{\rm c}(\omega)$ at T=20~K obtained with 
$\hbar\omega_{\rm i}=2.71$ eV (dots) and description (thin solid line). The background (dashed line) and 
phononic contributions ({\Bg}~mode: thick solid line, other phonons: dotted line) are also given. 
(b) Calculation of the cross-section with vanishing bare phonon intensity for the {\Bg} 
phonon (solid line) and comparison with the cross-section obtained with $\hbar\omega_{\rm i}=1.96$ eV (dots) 
at 20 K. Inset: Efficiencies of the Raman spectra
at 700 {\cm} (dots) in comparison with a calculation\cite{eff:cross} (solid line) according to ellipsometric data.\cite{Kircher:91}
\label{fig:spectra}}
\end{figure}

\begin{figure}
\caption{}{Bare (solid circles) and renormalized (open circles) frequency and 
linewidth of the $\rm B_{1g}$ phonon for spectra recorded with $\hbar\omega_{\rm i}=2.71$~eV.
Solid lines are fits to anharmonic decays for the data above $T_{\rm c}$ (dashed line). 
\label{fig:frequency}}
\end{figure}

\begin{figure}
\caption{}{Temperature dependence of the peak height $I_{2\Delta}$ 
and of the bare phonon intensity $I_{\rm pp}$ 
for $\hbar\omega_{\rm i}$=2.71, 2.41 and 1.96 eV. The dashed lines indicate $T_{\rm c}$.
\label{fig:intensity}}
\end{figure}


\begin{thebibliography}{1}
\bibitem{eff:cross}
The measured Raman intensity $I(\omega)$ and the efficiency $I_0(\omega)$
are related by $I(\omega)=[1+n(\omega)]I_0(\omega)=A[1+n(\omega)]{\rm Im}\chi(\omega)$ 
where $n(\omega)$ is the Bose factor and $A$ a proportionality constant 
due to the penetration depth and the complex refractive index.
\end{thebibliography}
\end{document}